# Advances in Numerical Simulation of High-Speed Impact Welding

Ali Nassiri [1*], Tim Abke [2], and Glenn Daehn [3]

[1]Simulation Innovation and Modeling Center (SIMCenter), The Ohio State University, Smith Laboratory Room 3076, 174 West 18th Avenue, Columbus, OH 43210, USA
[2]Honda R&D Americas, Inc., 21001 OH-739, Raymond, OH 43067, USA
[3] Department of Materials Science and Engineering, The Ohio State University, 116 W 19th Avenue, Room 141 Fontana Labs, Columbus, OH 43210, USA
*nassiri.3@osu.edu

## Abstract

Numerical simulations of high-speed forming and welding are of significant interest to industry, but are challenging due to the coupled physics and dynamic nature of the processes. With the advancement in hardware and computational capabilities, the next generation of computational methods, so called meshless methods, have received significant attention. Among all meshless methods, smoothed particle hydrodynamics (SPH) has received major consideration. The main advantage of the SPH method is to bypass the requirement for a numerical grid to calculate spatial derivatives. This avoids severe problems associated with mesh tangling and distortion which usually occur in Lagrangian analysis involving high-strain-rate events. In this study to better understand the effects of oxide layer, coating and diffused materials on weldability, a novel hybrid SPH platform was developed. Then, the high-speed impact between Steel/Steel, Copper/Titanium, and Aluminum/Steel were simulated. To experimentally validate the numerical efforts, results were compared to vaporizing foil actuator welding and explosive welding tests. Good agreement between the numerical simulations and experimental results provided confidence in the numerical modeling.

## 1. Introduction

In the automotive, aerospace and various other industries, there is a primary need for lightweight structures. This goal can be achieved by using multi-material assemblies. But owing to the disparate melting temperatures of the materials and concerns about brittle intermetallic compounds formation, traditional fusion welding processes cannot be used. One means to join dissimilar metals is through high-speed impact welding (HSIW). Typically, the HSIW process involves a high-speed, oblique collision between two metals arranged in relatively simple geometries. A striking characteristic of metals joined using HSIW is the emergence of a distinctive wavy pattern at the interface of two welded workpieces which is assumed to be necessary for creation of a quality impact weld (see Figure 1). To this end, numerical approaches are considered to be necessary, especially in predicting the shape and temperature distribution at the interface.

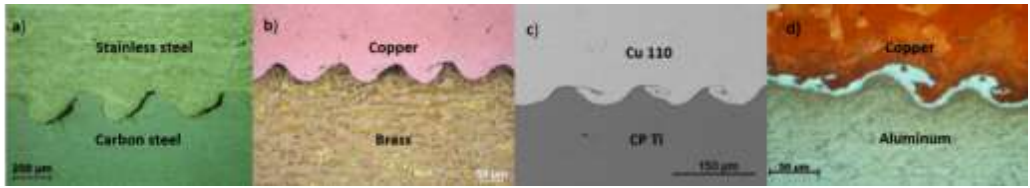

*Figure 1. A wavy morphology between a) Steel/Steel [1], Copper/Brass [2], C) Copper/Titanium [3], and d) Copper/Aluminum [4]*

However, finite element simulations of the HSIW process is very challenging, especially near the welding zone where large deformations occur. Thus, a traditional Lagrangian method where the mesh is fixed to the workpiece geometry fails as a result of excessive element distortion. Smoothed particle hydrodynamics (SPH) is a viable method to simulate the highly dynamic events. SPH is a Lagrangian technique where the coordinates move with the objects. The SPH particles are interpolation points from which values of functions and their derivatives can be estimated at discrete points in the continuum. The function values and their derivatives are found by a kernel approximation instead of being constructed from a grid. Since in SPH there is no connectivity between the particles, they are able to move relative to each other in the domain of simulation. Hence, metal jet emission can be simulated which was previously impossible using Lagrangian mesh-based numerical methods. Jetting is believed to be as a prerequisite for welding to occur. This phenomenon cleans nascent oxide layers, contaminants, and coating off the mating surfaces, and this allows solid state bonding without extensive melting and long range diffusion. In this study, in order to further understand the science behind the HSIW process and find a relationship between the composition of the jetted materials and weldability, a new hybrid computational platform was developed using a combination of SPH and traditional Lagrangian mesh-based methods. Then, the high-speed impact between Steel/Steel, Copper/Titanium, and Aluminum/Steel were simulated using this platform. To experimentally validate the numerical efforts, results were compared to vaporizing foil actuator welding (VFA) and explosive welding experimental tests.

## 2. Numerical Simulation Model

The numerical simulations were carried out in LS-DYNA software package. Figure 2a shows a schematic illustration of the traditional SPH oblique impact model. A flyer plate is inclined $\beta$ degree with respect to a target plate and given the initial impact velocity of $V_f$ through the flyer plate. To model a typical impact between two plates with the same size (e.g., a=c=2.5mm and b=d=2.5mm) over 2,000,000 particles with a size of 4µm needs to be generated. In the new proposed platform, SPH region was set to be 250µm in the thickness where the plates suffer heavy deformation by collision. A Lagrangian mesh-based platform was used for the rest of the model (See Figure 2b). The boundary between the SPH and Lagrangian regions was connected by defining a joint function. The joint function ties the two regions. In this model, the normal and shear stresses and their failure values are used as controlling parameters that define the strength of the tie. In the new platform less than 90,000 particles were generated. Another advantage of the hybrid platform is that different layers of materials can be added to the SPH section in which each layer has its own thermo-mechanical properties and particle size (see Figure 2c). In this study, the material behavior was modeled using the Johnson-Cook constitutive relationship. The pressure at the contact point, which is incorporated in the stress tensor, was computed using the Grüneisen state equation. Also, the experimentally measured impact velocity was used as an initial condition and input to the numerical code. See Ref [5,6] for more information about the governing equations and the material parameters.

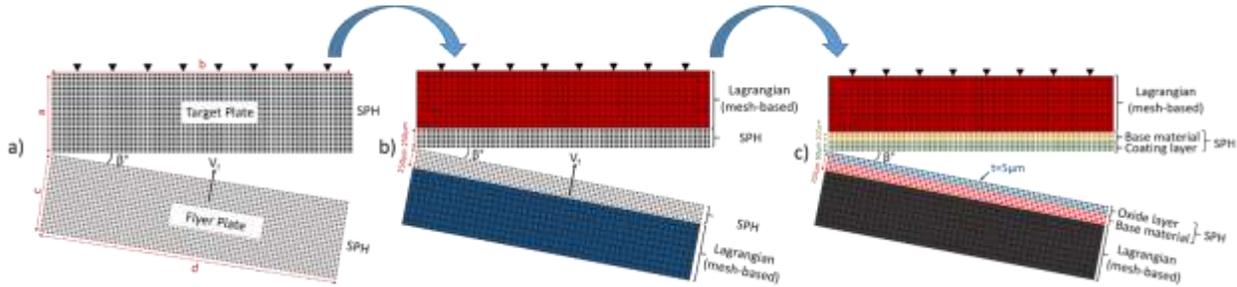

Figure 2. Schematic of the simulation setup; a) traditional SPH, b) hybrid platform, c) hybrid platform with coating and oxide layers

### 3. Results and Discussion

Figure 3 shows a feature-by-feature comparison between the numerical simulations and experimental tests in Steel/Steel and Copper/Titanium bimetallic systems. As is clear, the new platform was able to capture both wavy morphology and the swirling motion of the materials. Figure 4a shows the simulation result between an aluminum (Al6111-T4) plate impacting a dual phase high-strength steel (JAC980) plate.

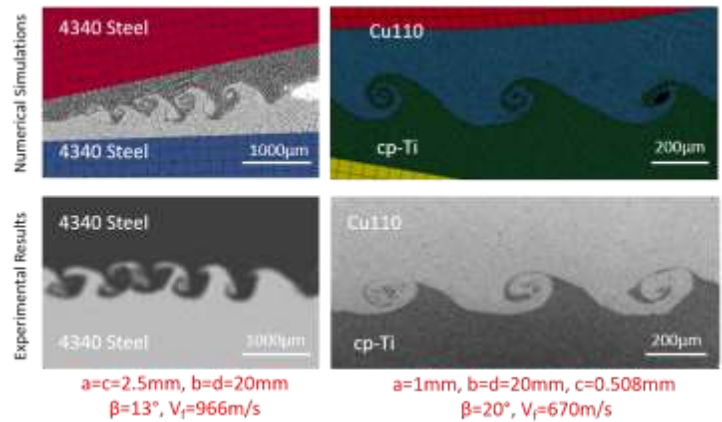

Figure 3. Comparison between the numerical simulations and experimental tests; Left column- impact between Steel/Steel using explosive welding [6]; right column- impact between Copper/Titanium using VFA

In JAC980, a galvanneal steel, to avoid corrosion, a layer of Zn coating is added to the base metal. As is shown, the hybrid platform was captured the jetted materials. The result shows that the ejected particles contain aluminum oxide, base aluminum, and zinc. Also, there is no indication of the base steel among the jetted materials which implies that pressure at the contact point was not enough to penetrate through the target plate and remove the coating. Therefore, no bond between the two base materials was cretaed. This can be confirmed by comparing the experimental result and numerical simulation at the end of the process which shows a failed welded sample and simulation, respectively (Figure 4b,c). In term of computational time, the suggested hybrid platforms are approximately 4 times faster than the traditional SPH platform.

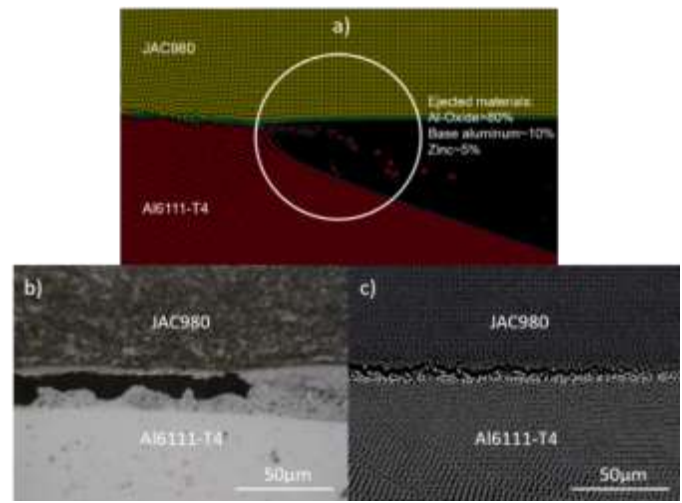

*Figure 4. a) Simulation result between Steel/Aluminum, b) failed sample, c) simulation result as the end of the process showing the interface is separated*

To validate the accuracy of the new platform, in the future, the composition of the jetted materials from the model will be compared with the jetted materials that will be collected in experiments by placing a witness block perpendicular to the direction of weld. The new platform can be utilized to reduce the number of trial-and-error iteration during the experimental tests by providing the numerically captured process parameters (e.g. impact angle and impact velocity).